# Leaky Integrate-and-Fire Mechanism in Exciton-Polariton Condensates for Photonic Spiking Neurons

K. Tyszka[1*], M. Furman[1], R.Mirek[1], M. Król[1], A. Opala[2], B. Seredyński[1], J. Suffczyński[1], W. Pacuski[1], M. Matuszewski[2], J. Szczytko[1] and B. Piętka[1†]
[1]*Institute of Experimental Physics, Faculty of Physics, University of Warsaw, ul. Pasteura 5, Warsaw PL-02-093, Poland*
[2]*Institute of Physics, Polish Academy of Sciences, Al. Lotników 32/46, Warsaw PL-02-668, Poland*

This paper introduces a new approach to neuromorphic photonics in which microcavities exhibiting strong exciton-photon interaction may serve as building blocks of optical spiking neurons. The experimental results demonstrate the intrinsic property of exciton-polaritons to resemble the Leaky Integrate-and-Fire spiking mechanism. It is shown that exciton-polariton microcavities when non-resonantly pumped with a pulsed laser exhibit leaky-integration due to relaxation of the excitonic reservoir, threshold-and-fire mechanism due to transition to Bose-Einstein condensate, and resetting due to stimulated emission of photons. These effects, evidenced in photoluminescence characteristics, arise within sub-ns timescales. The presented approach provides means for energy-efficient ultrafast processing of spike-like laser pulses at the level below 1 pJ/spike.

## 1 Introduction

Neuromorphic engineering aims to develop hardware capable of unconventional computing by emulating the physiology of the neuronal network of a brain.[1] Here, we particularly refer to Spiking Neural Networks (SNN) a special class of Artificial Neural Networks (ANN) often denoted as 3rd generation networks. [2] This notion reflects the promise of improvements in the computational power efficiency of SNN by maintaining a more strict analogy to brain-like processing with trains of asynchronous spikes. [3]

The concept of Neuromorphic Photonics introduced advantages of optical information processing into the neuromorphic engineering domain. [4] This especially addresses potentially limiting factors of more matured neuromorphic electronics. Although the progress within this domain is astonishing, most of the state-of-the-art processors are optimized for a specific goal, e.g. achieving low power consumption by utilizing digital representations of spiking signals,[5] providing high flexibility and reconfigurability based on von Neumann many-core architecture or high-speed processing based on analog neural circuits.[6,7] These approaches are a result of trade-offs between desirable objectives, being a consequence of fundamental limits related to electrical signals propagation.[8,9] For the same reason, such electronic systems rely on at most sub-μs timescales of operation to achieve asynchronous communication within a dense network of electronic interconnections.[10]

Optical implementations, on the other hand, may allow targeting sub-ns regimes with the gigahertz switching speeds simultaneously providing high communication bandwidth, and low cross-talk.[11] In connection with sub-ns pulsed lasers, photonics is very well-suited for ultrafast spike-based information processing requiring high interconnection densities.[12] It is expected that hypothetical integrated photonic spiking processors could potentially operate six orders of magnitude faster than neuromorphic electronics. [13]

Since the beginning, the engineering of SNN devices has focused on two mutually exclusive aspects, first to develop scalable, fast, and low-powered solutions, and second to faithfully model biological neurons.[8,14] The contradiction, as a rule of thumb, is that the more rich neuron models add more usefulness to neuromorphic hardware while being more computationally inefficient and harder to emulate in large networks. Within the neuromorphic electronics field, this issue has been often addressed, e.g. by introducing digital neuron representation,[15,16] optimizing spiking analog circuits, or the model itself to suit the hardware better.[17] The less mature optical domain research still seeks suitable solutions allowing the implementation of spiking neural networks.

In general, the minimum neuron functionality necessary to realize efficient and brain-like information processing is well approximated by the family of Integrate-and-Fire (IF) models.[18] The simplest IF neuron captures only the most basic biological neuron features, i.e. at least integration of input spikes and spike firing due to threshold crossing. Nevertheless, it has been demonstrated that networks of IF neurons are capable of visual pattern recognition,[19] saliency extraction,[20] speech recognition, or robot control.[21] Considerable effort was made to find physical phenomena within the all-optical domain which accurately mimics neuro-computational functionalities supported by IF models. One of the first reports has pointed to neuron-like pulse generation in a semiconductor resonator cavity with pump perturbation.[22] Later the excitability of semiconductor lasers has been often discussed in the context of neuronal excitability.[23,24] The turning point came after the first demonstration of fiber-based ultrafast Leaky-Integrate-and-Fire (LIF) neuron, a class of IF.[25] The semiconductor optical amplifier has been used to implement the LIF mechanism, although with periodic gain sampling. This work focused particularly on the optical realization of the neuronal model for neuromorphic computation. [26] Following this approach, within the last decade, there have been several reports pointing explicitly to similarities between spiking neurons and various optical effects. First reports focused on excitability in semiconductor ring lasers,[27] injection-locked Vertical Cavity Emitting Lasers (VCSEL),[28] and optically pumped VCSEL with the saturable absorber.[29] Soon later these systems were used to implement more advanced LIF functionalities,[30,31,32] or controlled generation of spiking patterns.[33] This was followed by demonstrating computational usefulness - temporal recognition tasks with chains of micro-lasers,[34] and pattern recognition based on coincidence detection of VCSEL spikes.[35] Recent advancements in integrated photonics also led to on-chip spiking neuron realizations with basic spiking neurons based on phase-change materials.[36] Although much progress has been made the optical approaches are much in their infancy in comparison to the electrical domain. Current efforts are focused on identifying the potential mechanisms for useful and flexible neuron implementation. Nevertheless, the foundations have been laid down proving the possibility of ultrafast neuromorphic processing. The building blocks of future optical SNN competitive to electronic solutions are yet to be clarified.

* ktyszka@fuw.edu.pl
† barbara.pietka@fuw.edu.pl

In this context, we propose a new solution in which exciton-polaritons (abbr. polaritons) in microcavities may provide building blocks for sub-ns efficient optical emulation of biological neurons. Polaritons are quasiparticles formed due to the strong interaction of photons confined in optical microcavity and excitons confined within embedded quantum wells. [37,38] Such light-matter coupling strengthens photon-photon interactions through the admixture of the excitonic component. This leads to remarkable phenomena like non-equilibrium Bose-Einstein condensation or superfluid-like states.[40] Importantly, polaritons also provide non-linear oscillatory dynamics under pulsed excitation.[41] These unique properties allowed the demonstration of Josephson junctions,[42] polariton transistors and logical gates,[43,44] classical artificial neurons,[45] non-linear phenomena at the femtojoule level,[46] and polariton-based reservoir computing.[47] Our proof-of-principle polariton binarized neural network has shown the capability of efficient handwritten digit recognition with high accuracy.[44] Moreover, we have shown that in general polariton-based networks could provide energy efficiency and performance density in inference tasks orders of magnitude higher than electronics.[48]

Here we report that exciton-polariton systems can mimic the Leaky-Integrate-and-Fire mechanism underlying most of the advanced IF spiking neuron models. Particularly, we experimentally show that polariton microcavities under non-resonant pulsed pumping exhibit spiking photoluminescence characteristics similar to the pulse response of a LIF neuron but within ps timescales. This effect arises due to the property of polaritons that can undergo a rapid transition to non-equilibrium Bose-Einstein condensate above the excitation threshold, followed by a strong spike in photoluminescence. We demonstrate this analogy in detail based on ultrafast measurements with a streak camera, focusing on microcavity pulse response to single or consecutive picosecond laser pulses. First, by comparing the LIF model with the polariton condensation mechanism we propose the polariton population as an internal state variable. This is equivalent to LIF membrane potential which governs the spiking behavior of the neuron. Then, step by step we designate other analogies to particular functionalities implied by the LIF model and experimentally demonstrate the physical mechanisms behind each of these similarities. This way we show that systems taking advantage of intrinsic properties of exciton-polaritons may become future building blocks of optical neurons and neuromorphic devices in general.

## 2 LIF neuron and its polariton analog

Within the LIF model, neuron dynamics are reproduced by the electrical circuit depicted in the inset of **Figure 1(a).** [49] The spikes of electrochemical signals (action potentials) are events idealistically

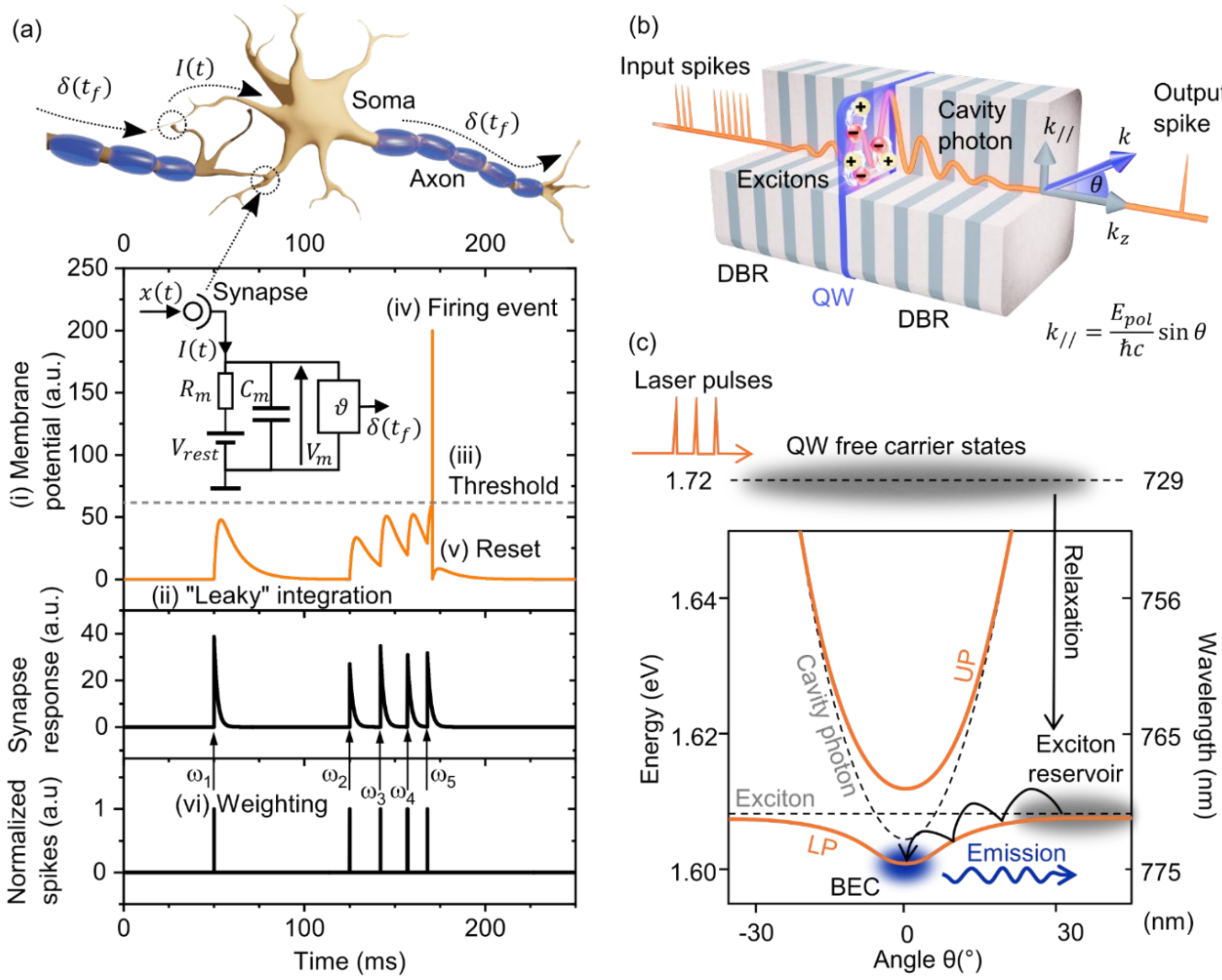

**Figure 1 (a)** Calculation of LIF spiking in response to multiple incoming spikes, ω – synapse efficacy. The main LIF neuron functionalities from (i) to (vi) are also depicted. Inset - LIF neuron circuit representation: *x(t)* - series of delta spikes from presynaptic neurons, *I(t)* - synapse output in a form of decaying current *Rm, Cm* - resistance and capacitance associated with the neuron's membrane. The RC circuit works as a leaky-integrator of membrane potential $V_m$ and drives the thresholder $\vartheta$. Threshold crossing leads to a firing event (spike emission). Here we assume $V_{reset} = V_{rest} = 0$. **(b)** Scheme of a polariton microcavity, its working principle, and relationship between the angle of microcavity far-field emission $\theta$ and in-plane momentum $k_{//}$. (c) Scheme of non-resonant optical excitation of polaritons, represented in energy vs angle $\theta$ of emission (equivalent to $k_{//}$), dashed lines - the pure cavity photon and quantum-well exciton dispersions, solid lines - the lower polariton (LP) branch (low energy side), and upper polariton (UP) branch (high energy side). Exemplary energies of the excitation pulse and emitted pulse are also depicted by (wavy arrows).

represented as delta functions $\delta(t - t_j^{(f)})$, where $t_j^{(f)}$ describes the moment of spike firing. The input from multiple presynaptic neurons is represented as a series of delayed spikes $x(t) = \sum_f \delta(t - t_j^{(f)})$ which are low-pass filtered at the synapse and generate the postsynaptic current $I(t) = \sum_j \omega_j \sum_f \alpha\left(t - t_j^{(f)}\right)$. The $\alpha$ function describes the synapse response to a delta spike and is discussed in more detail in the Methods section. The $\omega_j$ corresponds to the efficacy of connection of a neuron with j-th presynaptic neuron. The *I(t)* current either charges the capacitor or leaks through the resistor $R_m$ of a parallel *RC* circuit effectively causing the exponential decay of voltage across the capacitor plates. This potential difference represents the membrane potential of a neuron $V_m(t)$, and the *RC* circuit effectively works as a "leaky integrator" with a time constant $\tau_m = R_m C_m$. In result, the membrane potential decays in time according to the following Equation 1.

$$\tau_m \frac{dV_m(t)}{dt} = V_{rest} - V_m(t) + R_m I(t) \tag{1}$$

If $V_m(t)$ reaches the threshold before circuit discharge, a delta spike is emitted at the moment of threshold crossing $t^{(f)}$: $V_m(t^{(f)}) = \vartheta$. Immediately, the potential is reset to a new value $V_{reset}$. Here we assume $V_{reset} = V_{rest} = 0$, where $V_{rest}$ is the membrane resting potential. This corresponds to the case when the refractoriness mechanism is not implemented. We also assume that the modeled neuron responds only to excitatory inputs in the form of pulses.

This IF model instance reduces the biological neuron to six fundamental functionalities: (i) defines internal state variable representing the *membrane potential*, (ii) provides *leaky-integration* of input spikes which contributes to the build-up and decay of the membrane potential, (iii) sets the *threshold* level of membrane potential for realization of fire or no-fire decision, (iv) triggers the spike *firing event* which contributes to consecutive spike emission into the network, (v) forces the *reset* of the membrane potential. The LIF model also implies synapse functionality. i.e. (vi) weighted-summation of incoming spikes. According to this model, the sequence of LIF-like spiking in response to multiple incoming spikes is shown in **Figure 1(a)**.

Within the regime of pulsed non-resonant excitation of semiconductor microcavities in the strong light-matter coupling regime, we can derive the mechanism and phenomena which can reproduce six main LIF functionalities. We explain this analogy in the next sections.

## 2.1 Membrane potential and thresholding

Microcavity exciton-polaritons are bosonic quasiparticles formed due to the strong light-matter coupling of photons confined in optical microcavity and excitons typically confined within embedded quantum wells (QW), as depicted in **Figure 1(b)**. [50] Typical planar optical microcavities resemble a structure of an optical resonator composed of two facing Distributed Bragg Reflectors (DBR) separated by a sub-micron cavity (as shown in **Figure 1(b)**). The key condition required to obtain an exciton-polariton system is to achieve strong exciton-photon interaction by confining both photons and excitons into a small volume for an extended time. The former is provided by a high-quality microcavity, the latter by embedding an active medium into the cavity, e.g. semiconductor quantum well.

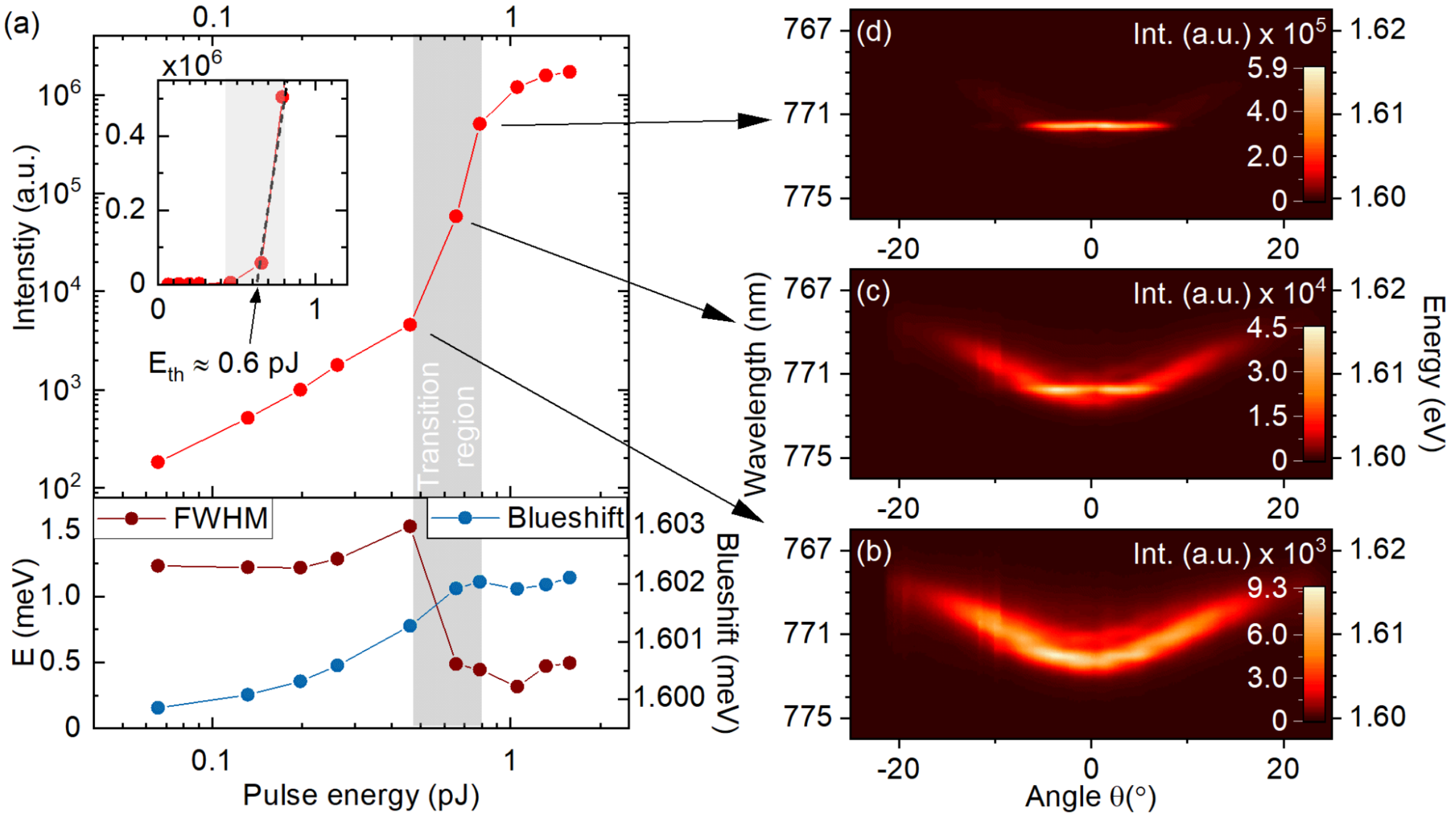

**Figure 2 (a)** Characteristics of time-averaged polariton microcavity photoluminescence vs laser pulse energy. Intensity thresholding occurs due to the transition of polaritons to a BEC (top). This is accompanied by emission line narrowing and energy blueshift at emission maximum occurring at zero angle (bottom). The pulse energy threshold level is extrapolated (inset). The grey area marks the BEC transition region. **(b, c, d)** The representative emission spectra accompanying the transition to BEC b) above the threshold, c) at the threshold, d) below the threshold.

The photon-exciton interaction leads to the creation of new eigenstates – upper and lower polaritons. Particularly, the polariton dispersion relation (the relation between polariton energy and in-plane momentum) takes the form of characteristic quasi-paraboloids and can be fully reproduced in the angular distribution of the microcavity far-field emission, as shown in **Figure 1(c).** [51] These states can be efficiently populated by particles through non-resonant laser pumping. Photo-excited free carriers, electrons and holes, relax through phonon scatterings, creating a source of incoherent excitons in the form of the *excitonic reservoir*. The reservoir serves as the source of excitons which may continuously refill the lower polariton states of the system driving the radiative relaxation of polaritons. For high enough laser pumping power, scattering becomes very effective, which leads to bosonic stimulation and generation of a polariton Bose-Einstein Condensate (BEC) accompanied by rapid relaxation of polaritons and strong microcavity emission.[39] The phenomenon is considered a non-equilibrium BEC due to the dissipative character of the system.[52] The short lifetime of polaritons renders polariton condensates out of equilibrium because the steady-state is a balance between losses and scattering from the excitonic reservoir.

The condensation mechanism can be directly evidenced by the observation of non-linear characteristics of time-averaged microcavity photoluminescence and emission spectra. Importantly the BEC transition is accompanied by characteristics that exhibit thresholding. We routinely observe such characteristics during experiments, as shown in **Figure 2(a).** Below a certain pumping power (power threshold) the response of the system can be described by a linear relationship. At this stage, polaritons are in a low-density regime, i.e. there are not enough polaritons in the system to form the condensate. The emission spectra reveal lower polariton quasi-parabolic dispersion (**Figure 2(b)**). If power increases above a certain level, the population of exciton-polaritons reaches a critical level (population threshold level) for condensation to occur (**Figure 2(c)**). A superlinear increase in the intensity is observed due to collective and coherent light emission from BEC (**Figure 2(d)**). This effect is accompanied by other characteristic properties, narrowing of the emission line and emission blueshift, (**Figure 2(a)**). Here, typically BEC transition occurs at excitation pulse energies below 1 pJ. Due to the finite size of the system, the transition is not sharp. The BEC transition region has a width of approx. 0.3 pJ in the function of pulse energy. Some emission occurs regardless of whether the *BEC threshold* was reached, but this is weak and can be cut off by spectral and spatial filters. It is expected that this kind of modification may improve the overall performance of the polariton microcavity-based thresholder. In our optical LIF mechanism proposal, the phenomenon of *BEC* transition under non-resonant pumping underlies the thresholding mechanism (iii). The choice of this non-linear phenomenon implies the *polariton population* as an analog of (i) membrane potential, as the internal state variable.

## 2.2 Threshold-and-fire mechanism and reset

Inherently, under pulsed non-resonant excitation, the BEC threshold crossing is accompanied by pulse generation due to stimulated polariton radiative emission. Intending to implement the neuron firing event (iv) after threshold crossing (iii) we investigated the response of microcavity polaritons to picosecond non-resonant pulse excitation. By high-resolution time-resolved measurements with a streak camera, we observed the sub-ns response of the polariton system as shown in **Figure 3(a).**

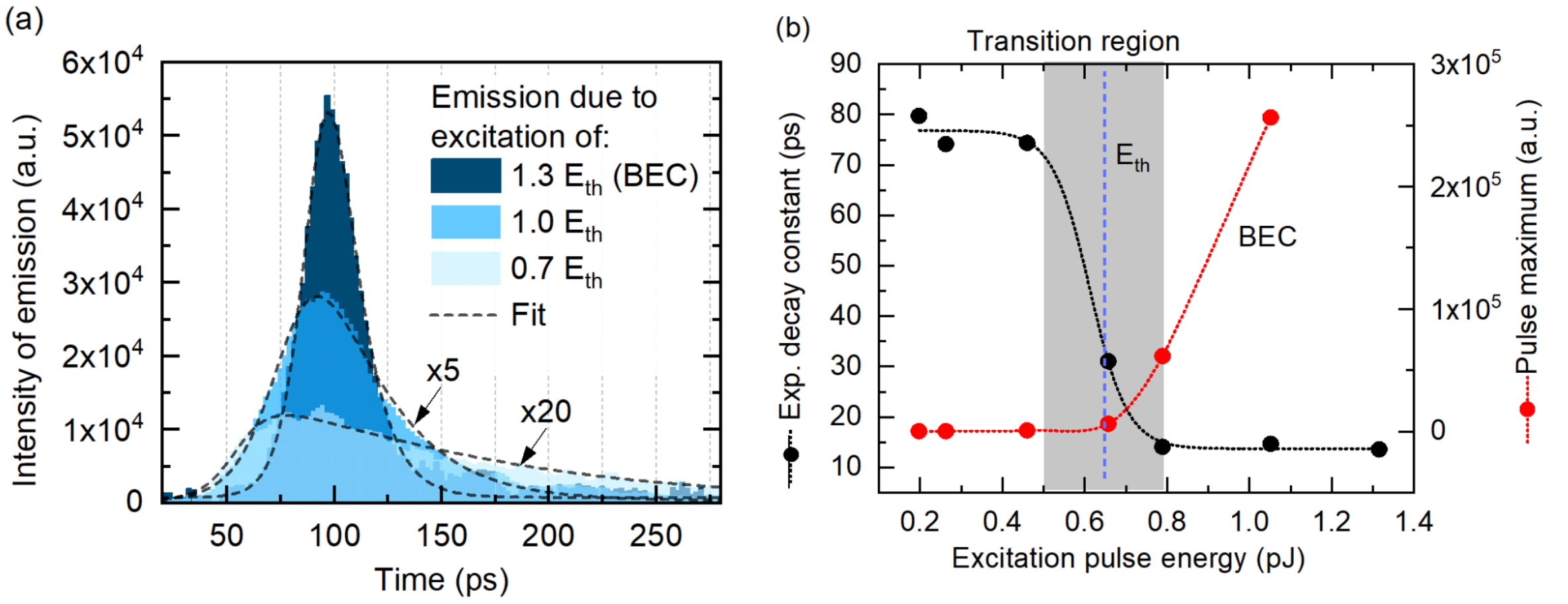

**Figure 3 (a)** High-resolution time-resolved characteristics of emission pulses obtained with streak camera results. The energies of excitation pulses correspond to the limits of the BEC transition region (gray area (b)). For clarity, emissions due to lower energies are multiplied by 20 and 5 respectively. Dashed lines mark exponential decay interpolation. **(b)** Emission pulses intensity maxima (red) and decay times (black) vs excitation pulse energies (black).

The incoming laser pulse generates free carriers which rapidly, i.e. in sub-ns scale, scatter to the lower energy states and create the excitonic reservoir which serves as the source of excitons. The reservoir lifetime is a property of the material used to form microcavity quantum wells and exceeds the lifetime of photons trapped in a microcavity. Below the BEC threshold, this is evidenced by immediate but extended in time emission of photons initially bound to the lower polariton branch (**Figure 3(a)).** In this regime, in the first approximation, the response decay time corresponds to the reservoir lifetime, here of the order of 100 ps. The amplitude of response inherently depends on the excitation power as more polaritons render an increased relaxation rate. As higher energy laser pulses generate more polaritons the density is increased and above critical value transition to BEC occurs. This is accompanied by a slight delay of pulse emission, pulse narrowing, and sharp decay due to rapid polariton relaxation and emission under bosonic stimulation (**Figure 3(b)**). The polariton decay time above the condensation threshold becomes short, and here is on the order of 10 ps. The decaying emissions presented here reflect changes in the polariton density.

Concerning the neuron functionalities, our measurements show that polariton microcavity under pulsed excitation regime works as an integrating and leaky element (ii). For polariton densities below the BEC threshold emits weak and slowly decaying pulses. For polariton densities above the BEC provides a strong and narrow pulse response. Importantly, above the BEC threshold, the polariton relaxation leads to a rapid decrease in the polariton population and depletion of the excitonic reservoir. This corresponds to a LIF-like mechanism when threshold crossing (iii) due to a build-up of membrane potential leads to subsequent spike firing event (vi) and reset (v), while neuron stimulation below threshold renders only membrane potential decay due to leaky-integration (ii). Due to sub-ns time scales of adapted physical phenomena, this approach provides ultrafast processing capabilities with the energy efficiency of approx. 0.6 pJ/spike.

## 2.3 Leaky integration of consecutive pulses and reset

The investigation of the microcavity pulse response confirms the resemblance of the LIF mechanism. Here, we address the evidence of proper processing of consecutive spikes, especially at the condition when spikes may collectively induce threshold crossing. To confirm this correspondence with the LIF model we investigated microcavity response to consecutive laser pulses. We also compared these results to the LIF model by fitting the model parameters. Here, we assumed a more realistic gaussian form of spikes instead of delta functions (see Equation 4 in Methods). Results are shown in **Figure 4.** For reference, **Figure 4(a, b)** also shows time-resolved emission due to excitation with each pulse separately. In comparison, the emission due to the second pulse is stronger in the case of excitation with consecutive pulses. This is because the excitonic reservoir generated by the first excitation is replenished by the following one. This is possible when the interpulse interval is shorter than the reservoir lifetime. In this case, the reservoir lifetime is also extended. The excitons remaining from the prior excitation contribute to the increased polariton density and a stronger secondary emission. This is analogical to the membrane potential (i) build-up due to leaky-integration (ii) of incoming spikes in the numerical simulations of the LIF model (1), as depicted in **Figure 4 (c).** Importantly, in the experiment, we used laser pulses with different energies to realize excitation with weighted pulses. This is equivalent to feeding the LIF neuron through synapses with different efficacies. The result confirms that our optical analog of the LIF mechanism is capable of processing weighted spikes and different temporal coding schemes.

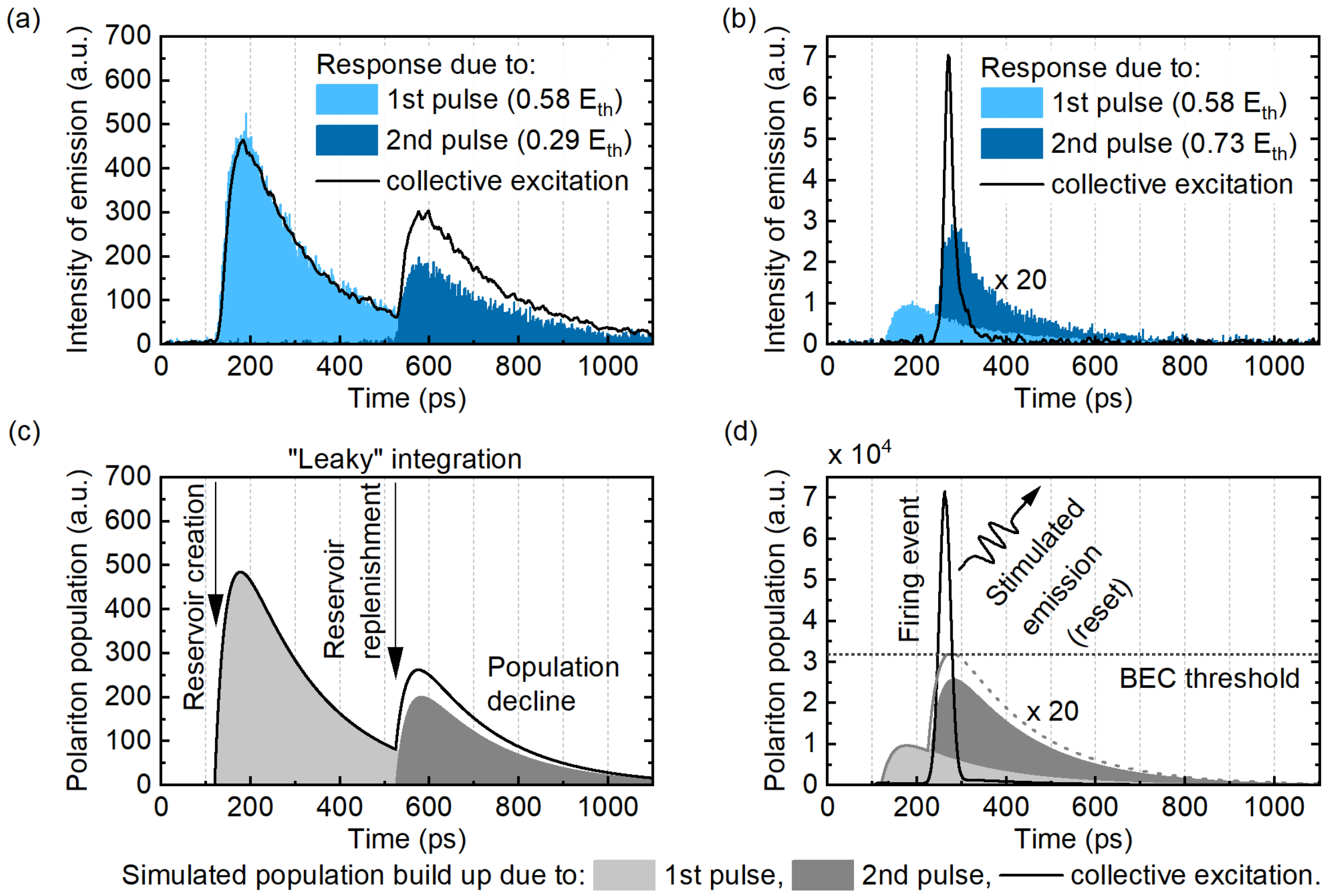

**Figure 4** Top - polariton microcavity response to excitation with two consecutive laser pulses. (a) Pulses do not induce BEC. (b) Pulses collectively induce BEC and strong pulse emission, single pulses are multiplied by a factor of 20 for clarity. Bottom – simulated polariton population build-up, calculated using the LIF model (1). (c) The case analogous to the LIF membrane potential build-up below the threshold. (d) The case analogous to LIF threshold-and-fire mechanism with reset, single pulses are multiplied by a factor of 20 for clarity.

When consecutive pulses collectively induce build-up of polariton density, it may reach the BEC threshold level as in **Figure 4(b)**. The transition leads to a stronger secondary emission with a shorter lifetime, a faster decrease of the polariton population, and faster reservoir depletion. This results in the emission of a well-defined, sharp output spike, and confirms the analogy to the threshold-and-fire mechanism (iii, iv) induced by consecutive spikes stimulating the LIF neuron, as depicted in **Figure 4(d)**. Moreover, the observed faster reservoir depletion even after excitation with consecutive pulses confirms that the resetting mechanism (v) is present here.

## 3 Conclusions

Our experimental results confirm that in the regime of pulsed non-resonant excitation of exciton-polaritons, the semiconductor microcavities reproduce the most fundamental spiking neuron functionalities imposed by the LIF model. By combining the effects of (i) potential-like build-up and (ii) leaky-integration due to generation of the excitonic reservoir and polariton accumulation, (iii) thresholding due to BEC, (vi) subsequent pulse emission due to rapid relaxation of the BEC, and (v) resetting due to reservoir depletion, it is possible to develop an optical analog of LIF neuron capable of processing asynchronous spikes represented by picosecond laser pulses. Importantly, due to sub-ns times scales of adapted physical phenomena, our approach provides means for energy-efficient ultrafast processing capabilities at the level below 1 pJ/spike in comparison to LIF electronic realizations operating at megahertz regimes with tens of pJ/spike. [53]

We have demonstrated that polariton microcavities can serve as a building block of future optical neurons by providing the LIF mechanism. The LIF model as a mathematical concept does not have to cope with network connectivity as it is implemented at the level of the algorithm. In hardware realization however, networkability is a key factor, and here it is not provided. This is a more general problem of many photonic computing realizations, i.e. meeting the qualitative scalability criteria required by any complex system of nodes capable of computing (e.g. cascadability, logic-level restoration, fan-in).[54] Works in polaritonics show that it is possible to overcome these limitations by resonant or mixed-type pumping.[47,56,57] Based on recent efficiency analyzes of polariton networks we can expect that this will not drastically increase the energy requirements per spike.[48]

In addition to the scalability criterion, the network of nodes has to provide functionality specific for SNN, i.e. neuron input weighting and summation, and capability of configuring the weights in the process of learning. We showed that our LIF optical analog is capable of processing weighted spikes. The problem of implementing optical weights can be solved using various methods, as shown in a number of works. [57-61]

It has been argued that simple IF models despite their advantageous simplicity do not produce a rich enough range of behaviors thus compromising the usefulness of hardware implementations.[62] We believe that by taking advantage of a rich repertoire of remarkable phenomena existing in polariton systems further analogies to more advanced neuron functionalities may be demonstrated. This promise is even more intriguing considering that the polaritonic platform potentially provides means for ultrafast and energy efficient spike processing which copes with aspirations of the neuromorphic engineering field.

## 4 Methods

### *Sample*

The sample used is a semiconductor heterostructure with two distributed Bragg reflectors with 16 and 19 alternating (Cd,Zn,Mg)Te/(Cd,Mg)Te layers separated by 600 nm thick (Cd,Zn,Mg)Te layer. The microcavity has a quality factor of 300 and contains three pairs of quantum wells of 20 nm with a small concentration (about 0.5%) of manganese ions each. The structure was grown on a (100)-oriented GaAs substrate by molecular beam epitaxy. The detailed scheme of the structure is included in Supplementary Information (SI) as **Figure S1**.

### *Experimental setup*

The experimental setup is depicted in **Figure S2** in SI. The sample was placed in a chamber of a confocal microscope and cooled down to 4.2 K with liquid helium. A picosecond Ti:sapphire laser with 76MHz repetition rate to create approx. 3 ps laser pulses of σ+ polarized light at the energy $E_{exc}$ = 1.724 eV ($\lambda_{exc}$ = 719 nm). First, to generate two consecutive pulses the pulsed laser beam was split in two. One of the beams was delayed with a free-space tunable delay line. The power of each pulse was tuned with variable neutral-density filters. The polarization was controlled with a set of waveplates. Later, two pulses were combined with a beam splitter. This pulsed beam was focused on the sample by objective with a high numerical aperture (of 0.68). The non-resonant excitation induced pulsed microcavity emission due to polariton generation. The emission was collected with the same objective. The detection optical setup consisted of a spectrometer, CCD camera, power meter, and streak camera. The emission could be resolved in real space or reciprocal space.

### *LIF simulation*

The numerical simulation procedure follows the LIF model described in detail in Reference 43. The model describes stimulation by synaptic currents generated by a synapse in response to presynaptic delta spikes. This response is described by $\alpha$-function and takes the form of:

$$\alpha(s) = \frac{q}{\tau_s} exp\left(-\frac{s}{\tau_s}\right) \Theta(s) \quad (3)$$

where $q$ is the total charge injected via the synapse, and $\Theta(s)$ is the Heaviside step function. We used Equation 1, 2, and 3 to fit the LIF response to our experimental results. Instead of representing spikes as $\delta$-functions we used Gaussian pulses according to Equation 4, to simulate light pulses typical for experiments. The $t_j^{(f)}$ describes the moment of incoming of f-th spike from j-th presynaptic neuron.

$$G(t) = a\, exp\left(-\frac{\left(t-t_j^{(f)}\right)^2}{c^2}\right) \quad (4)$$

**Supporting Information**

Supporting Information is available from the Wiley Online Library or the author.

**Acknowledgments**
K.T. acknowledges support from National Science Center, Poland Grant No. 2020/04/X/ST7/01379. B.P. acknowledges support from National Science Center, Poland Grant 2020/37/B/ST3/01657. R.M. acknowledges support from National Science Center, Poland Grant 2019/33/N/ST3/02019

## References

[1] G. Indiveri and T. K. Horiuchi, *Frontiers in Neuroscience* **2011**, *5*, 118.

[2] W. Maass, and C. M. Bishop (eds.), Pulsed Neural Networks (MIT Press, Cambridge, Mass, 1999).

[3] W. Maass, *Neural Networks* **1997**, *10*, 1659–1671.

[4] P. R. PVrucnal, B. J. Shastri, and M. C. Teich, Neuromorphic Photonics (CRC Press, Boca Raton, 2017).

[5] F. Akopyan, J. Sawada, A. Cassidy, R. Alvarez-Icaza, J. Arthur, P. Merolla, N. Imam, Y. Nakamura, P. Datta, G.-J. Nam, B. Taba, M. Beakes, B. Brezzo, J. B. Kuang, R. Manohar, W. P. Risk, B. Jackson, and D. S. Modha, *IEEE Transactions on Computer-Aided Design of Integrated Circuits and Systems* **2015**, *34*, 1537–1557.

[6] S. Furber, and A. Brown, in: 2009 Ninth International Conference on Application of Concurrency to System Design: (2009), pp. 3–12.

[7] S. Scholze, S. Schiefer, J. Partzsch, S. Hartmann, C. Mayr, S. Höppner, H. Eisenreich, S. Henker, B. Vogginger, and R. Schüffny, *Frontiers in Neuroscience* **2011**, *5*, 117.

[8] D. A. B. Miller, *Proceedings of the IEEE* **2000**, *88*, 728–749.

[9] J. Hasler, and H. Marr, *Frontiers in Neuroscience* **2013**, *7*, 118.

[10] A. R. Young, M. E. Dean, J. S. Plank, and G. S. Rose, *IEEE Access* **2019**, *7*, 135606–135620.

[11] B. J. Shastri, A. N. Tait, T. F. de Lima, M. A. Nahmias, H.-T. Peng, and P. R. Prucnal, *Encyclopedia of Complexity and Systems Science* **2018**, 1–37.

[12] P. R. Prucnal, B. J. Shastri, T. Ferreira, D. Lima, M. A. Nahmias, and A. N. Tait, *Adv. Opt. Photon.* **2016** , *8*, 228–299.

[13] T. F. de Lima, B. J. Shastri, A. N. Tait, M. A. Nahmias, and P. R. Prucnal, *Nanophotonics* **2017**, *6*, 577–599.

[14] K. Roy, A. Jaiswal, and P. Panda, *Nature* **2019**, *575*, 607–617.

[15] A. S. Cassidy, P. Merolla, J. V. Arthur, S. K. Esser, B. Jackson, R. Alvarez-Icaza, P. Datta, J. Sawada, T. M. Wong, V. Feldman, A. Amir, D. B.-D. Rubin, F. Akopyan, E. McQuinn, W. P. Risk, and D. S. Modha, in: The 2013 International Joint Conference on Neural Networks (IJCNN): (IEEE, Dallas, TX, USA, 2013), pp. 1–10.

[16] K. A. Boahen, *IEEE Transactions on Circuits and Systems II: Analog and Digital Signal Processing* **2000**, *47*, 416–434.

[17] P. Livi, and G. Indiveri, in: 2009 IEEE International Symposium on Circuits and Systems: (2009), pp. 2898–2901.

[18] A. N. Burkitt, *Biol Cybern* **2006**, *95*, 1–19.

[19] S. G. Wysoski, L. Benuskova, and N. Kasabov, *Neurocomputing* **2008**, *71*, 2563–2575.

[20] T. Masquelier, and S. J. Thorpe, *PLOS Computational Biology* **2007**, *3*, e31.

[21] H. Paugam-Moisy, and S. Bohte, in: Handbook of Natural Computing, G. Rozenberg, T. Bäck, and J. N. Kok (eds.): (Springer Berlin Heidelberg, Berlin, Heidelberg, 2012), pp. 335–376.

[22] A. V. Grigor'yants, and I. N. Dyuzhikov, *Quantum Electron.* **1994**, *24*, 469–470.

[23] H. J. Wünsche, O. Brox, M. Radziunas, and F. Henneberger, *Phys. Rev. Lett.* **2001**, *88*, 023901.

[24] B. Krauskopf, K. Schneider, J. Sieber, S. Wieczorek, and M. Wolfrum, *Optics Communications* **2003**, *215*, 367–379.

[25] D. Rosenbluth, K. Kravtsov, M. P. Fok, and P. R. Prucnal, *Opt. Express, OE* **2009**, *17*, 22767–22772.

[26] K. Kravtsov, M. P. Fok, D. Rosenbluth, and P. R. Prucnal, *Opt. Express, OE* **2011**, *19*, 2133–2147.

[27] W. Coomans, L. Gelens, S. Beri, J. Danckaert, and G. Van der Sande, *Phys. Rev. E* **2011**, *84*, 036209.

[28] A. Hurtado, K. Schires, I. D. Henning, and M. J. Adams, *Appl. Phys. Lett.* **2012**, *100*, 103703.

[29] S. Barbay, R. Kuszelewicz, and A. M. Yacomotti, *Opt. Lett.* **2011**, *36*, 4476.

[30] F. Selmi, R. Braive, G. Beaudoin, I. Sagnes, R. Kuszelewicz, T. Erneux, and S. Barbay, *Phys. Rev. E* **2016**, *94*, 042219.

[31] F. Selmi, R. Braive, G. Beaudoin, I. Sagnes, R. Kuszelewicz, and S. Barbay, *Phys. Rev. Lett.* **2014**, *112*, 183902.

[32] F. Selmi, R. Braive, G. Beaudoin, I. Sagnes, R. Kuszelewicz, and S. Barbay, *Phys. Rev. Lett.* **2014**, *112*, 183902.

[33] A. Hurtado, and J. Javaloyes, *Appl. Phys. Lett.* **2015**, *107*, 241103.

[34] V. A. Pammi, K. Alfaro-Bittner, M. Clerc, and S. Barbay, *IEEE Journal of Selected Topics in Quantum Electronics* **2020**, *26*, 1–7.

[35] J. Robertson, M. Hejda, J. Bueno, and A. Hurtado, *Sci Rep* **2020**, *10*, 6098.

[36] J. Feldmann, N. Youngblood, C. D. Wright, H. Bhaskaran, and W. H. P. Pernice, *Nature* **2019**, *569*, 208–214.

[37] I. Carusotto, and C. Ciuti, *Rev. Mod. Phys.* **2013**, *85*, 299–366.

[38] A. Kavokin, and G. Malpuech, Cavity Polaritons, 1. ed (Elsevier, Acad. Press, Amsterdam, 2003).

[39] J. Kasprzak, M. Richard, S. Kundermann, A. Baas, P. Jeambrun, J. M. J. Keeling, F. M. Marchetti, M. H. Szymańska, R. André, J. L. Staehli, V. Savona, P. B. Littlewood, B. Deveaud, and L. S. Dang, *Nature* **2006**, *443*, 409–414.

[40] R. T. Juggins, J. Keeling, and M. H. Szymańska, *Nat Commun* **2018**, *9*, 4062.

[41] A. Opala, M. Pieczarka, and M. Matuszewski, *Phys. Rev. B* **2018**, *98*, 195312.

[42] M. Abbarchi, A. Amo, V. G. Sala, D. D. Solnyshkov, H. Flayac, L. Ferrier, I. Sagnes, E. Galopin, A. Lemaître, G. Malpuech, and J. Bloch, *Nature Phys* **2013**, *9*, 275–279.

[43] D. Ballarini, M. De Giorgi, E. Cancellieri, R. Houdré, E. Giacobino, R. Cingolani, A. Bramati, G. Gigli, and D. Sanvitto, *Nat Commun* **2013**, *4*, 1778.

[44] R. Mirek, A. Opala, P. Comaron, M. Furman, M. Król, K. Tyszka, B. Seredyński, D. Ballarini, D. Sanvitto, T. C. H. Liew, W. Pacuski, J. Suffczyński, J. Szczytko, M. Matuszewski, and B. Piętka, *Nano Lett.* **2021**, *21*, 3715–3720.

[45] T. C. H. Liew, A. V. Kavokin, and I. A. Shelykh, *Phys. Rev. Lett.* **2008**, *101*, 016402.

[46] A. Dreismann, H. Ohadi, Y. del Valle-Inclan Redondo, R. Balili, Y. G. Rubo, S. I. Tsintzos, G. Deligeorgis, Z. Hatzopoulos, P. G. Savvidis, and J. J. Baumberg, *Nature Mater* **2016**, *15*, 1074–1078.

[47] D. Ballarini, A. Gianfrate, R. Panico, A. Opala, S. Ghosh, L. Dominici, V. Ardizzone, M. De Giorgi, G. Lerario, G. Gigli, T. C. H. Liew, M. Matuszewski, and D. Sanvitto, *Nano Lett.* **2020**.

[48] M. Matuszewski, A. Opala, R. Mirek, M. Furman, M. Król, K. Tyszka, T. C. H. Liew, D. Ballarini, D. Sanvitto, J. Szczytko, and B. Piętka, *Phys. Rev. Applied* **2021**, *16*, 024045.

[49] W. Gerstner, and W. M. Kistler, Spiking Neuron Models: Single Neurons, Populations, Plasticity (Cambridge University Press, Cambridge, 2002).

[50] A. Kavokin (ed.), Microcavities, Second Edition (Oxford University Press, Oxford ; New York, NY, 2017).

[51] R. Houdré, C. Weisbuch, R. P. Stanley, U. Oesterle, P. Pellandini, and M. Ilegems, *Phys. Rev. Lett.* **1994**, *73*, 2043–2046.

[52] M. H. Szymańska, J. Keeling, and P. B. Littlewood, *Phys. Rev. Lett.* **2006**, *96*, 230602.

[53] S. Dutta, V. Kumar, A. Shukla, N. R. Mohapatra, and U. Ganguly, *Sci Rep* **2017**, *7*, 8257.

[54] D. A. B. Miller, *Nature Photonics* **2010**, *4*, 3–5.

[55] A. V. Zasedatelev, A. V. Baranikov, D. Urbonas, F. Scafirimuto, U. Scherf, T. Stöferle, R. F. Mahrt, and P. G. Lagoudakis, *Nat. Photonics* **2019**, *13*, 378–383.

[56] A. V. Zasedatelev, A. V. Baranikov, D. Sannikov, D. Urbonas, F. Scafirimuto, V. Y. Shishkov, E. S. Andrianov, Y. E. Lozovik, U. Scherf, T. Stöferle, R. F. Mahrt, and P. G. Lagoudakis, *Nature* **2021**, *597*, 493–497.

[57] J. W. Goodman, A. R. Dias, and L. M. Woody, *Opt. Lett.* **1978**, *2*, 1.

[58] Y. Shen, N. C. Harris, S. Skirlo, M. Prabhu, T. Baehr-Jones, M. Hochberg, X. Sun, S. Zhao, H. Larochelle, D. Englund, and M. Soljacic, *Nature Photon* **2017**, *11*, 441–446.

[59] J. Spall, X. Guo, X. Guo, T. D. Barrett, A. I. Lvovsky, and A. I. Lvovsky, *Opt. Lett., OL* **2020**, *45*, 5752–5755.

[60] J. Feldmann, N. Youngblood, M. Karpov, H. Gehring, X. Li, M. Stappers, M. Le Gallo, X. Fu, A. Lukashchuk, A. S. Raja, J. Liu, C. D. Wright, A. Sebastian, T. J. Kippenberg, W. H. P. Pernice, and H. Bhaskaran, *Nature* **2021**, *589*, 52–58.

[61] T. Zhou, X. Lin, J. Wu, Y. Chen, H. Xie, Y. Li, J. Fan, H. Wu, L. Fang, and Q. Dai, *Nat. Photonics* **2021**, *15*, 367–373.

[62] E. M. Izhikevich, *IEEE Trans. Neural Netw.* **2004**, *15*, 1063–1070.

# Leaky integrate-and-fire mechanism in exciton-polariton condensates for photonic spiking neurons

Supporting Information

K. Tyszka*, M. Furman, R. Mirek, M. Król, B. Seredyński, J. Suffczyński, W. Pacuski, J. Szczytko and B. Piętka*

E-mail: ktyszka@fuw.edu.pl, barbara.pietka@fuw.edu.pl

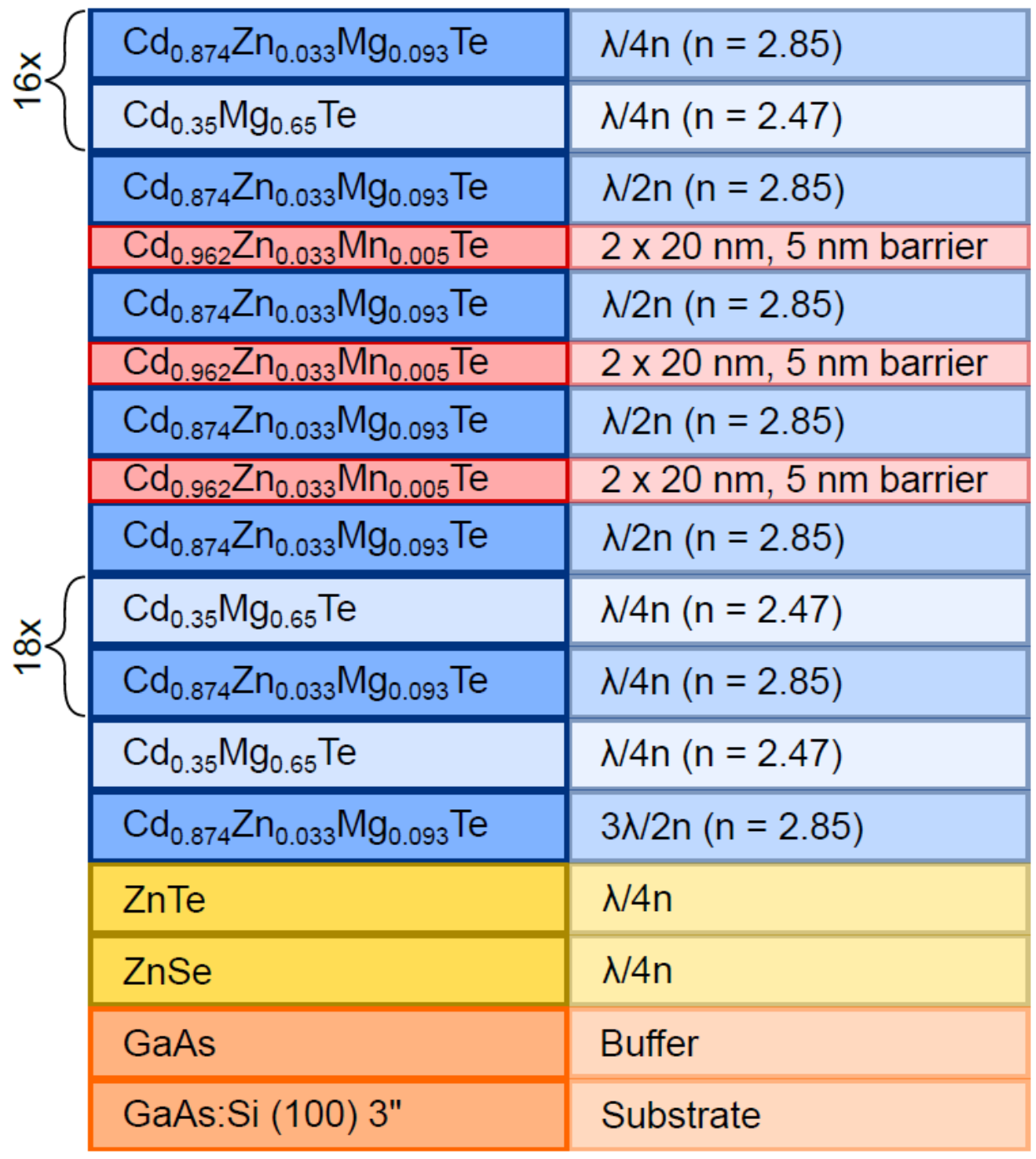

**Figure S1.** The structure of the sample: six 20 nm-wide QWs (red) containing manganese ions placed between 19 layers (bottom) and 16 layers (top) of non-magnetic distributed Bragg reflectors (alternating light blue and blue layers).

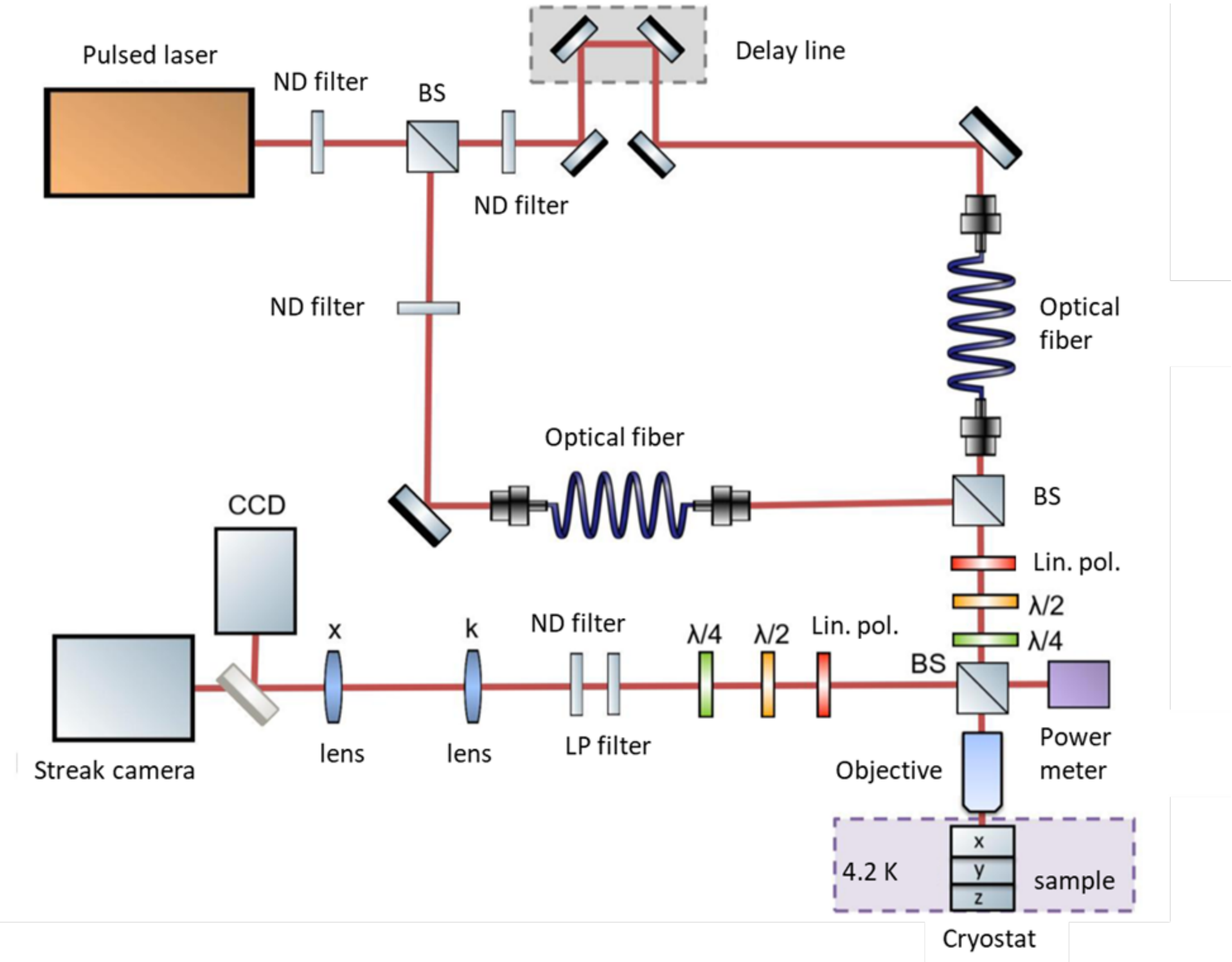

**Figure S2.** The experimental setup. The sample was placed in a chamber of a confocal microscope and cooled down to 4.2 K with liquid helium. A picosecond Ti:sapphire laser with 76MHz repetition rate to create approx. 3 ps laser pulses of σ+ polarized light at the energy $E_{exc} = 1.724$ eV ($\lambda_{exc} = 719$ nm). First, to generate two consecutive pulses the pulsed laser beam was split in two. One of the beams was delayed with a free-space tunable delay line. The power of each pulse was tuned with variable neutral-density filters. The polarization was controlled with a set of waveplates. Later, two pulses were combined with a beam splitter. This pulsed beam was focused on the sample by objective with a high numerical aperture (of 0.68). The non-resonant excitation induced pulsed microcavity emission due to polariton generation. The emission was collected with the same objective. The detection optical setup consisted of a spectrometer, CCD camera, power meter, and streak camera. The emission could be resolved in real space or reciprocal space.